\documentclass[conference,twocolumn,twoside,a4paper]{IEEEtran} 
\usepackage{adjustbox}
\usepackage{cite}
\usepackage{mathtools}
\usepackage{amsthm}
\usepackage{graphicx}
\usepackage[caption=false,font=footnotesize]{subfig}
\usepackage{paralist,lipsum}
\usepackage{paralist,lipsum}
\usepackage{amsmath,amsfonts,amssymb,stackrel}
\usepackage{verbatim}
\usepackage{enumerate}                     
\usepackage{psfrag}
\usepackage[nolist]{acronym}  
\usepackage[utf8]{inputenc}
\usepackage{tikz}
\usepackage{circuitikz}
\usetikzlibrary{shapes.geometric,arrows,positioning,patterns,matrix}
\usepackage{pgfplots}
\usepackage{pgfplotstable}
\usepgfplotslibrary{patchplots}
\pgfplotsset{compat=1.13}
\usepackage{tabstackengine} 
\usepackage{bm}
\usepackage{cuted}
\usepackage{flushend}
\usepackage[linesnumbered,ruled]{algorithm2e}

\def\Nj{N_\text{J}}
\def\Nt{N_\text{T}}
\def\Nr{N_\text{R}}

\def\x{\mathbf{x}}
\def\y{\mathbf{y}}

\newcommand{\trans}[1]{{#1}^{\mathsf{T}}}                     	

\begin{acronym}
\scriptsize
\acro{AD}{anomaly detection}
\acro{ADC}{analog to digital converter}
\acro{AE}{autoencoder}
\acro{AoA}{angle of arrival}
\acro{AoD}{angle of departure}
\acro{AWGN}{additive white Gaussian noise}
\acro{BS}{base station}
\acro{CRS}{cell-specific reference signal}
\acro{PDSCH-DMRS}{physical downlink shared channel-demodulation reference signal}
\acro{CDF}{cumulative distribution function}
\acro{c.d.f.}{cumulative distribution function}
\acro{CR}{cognitive radio}
\acro{DAA}{detect and avoid}
\acro{DFT}{discrete Fourier transform}
\acro{DRFM}{digital radio frequency memory}
\acro{DVB-T}{digital video broadcasting\,--\,terrestrial}
\acro{ECC}{European Community Commission}
\acro{ECCM}{electronic counter-countermeasures}
\acro{ECM}{electronic countermeasure}
\acro{EIRP}{effective isotropic radiated power}
\acro{FCC}{Federal Communications Commission}
\acro{FFT}{fast Fourier transform}
\acro{GPS}{Global Positioning System}
\acro{i.i.d.}{independent, identically distributed}
\acro{IFFT}{inverse fast Fourier transform}
\acro{IoT}{internet of things}
\acro{ISAC}{ integrated sensing and communication}
\acro{ITU}{International Telecommunication Union}
\acro{JSAC}{joint sensing and communication}
\acro{LOS}{line-of-sight}
\acro{MB}{multiband}
\acro{MC}{multicarrier}
\acro{MF}{matched filter}
\acro{MI}{mutual information}
\acro{MIMO}{multiple-input multiple-output}
\acro{MWS}{multi-look wideband sensing}
\acro{MRC}{maximal ratio combining}
\acro{MMSE}{minimum mean-square error}
\acro{MSE}{mean-square error}
\acro{NLOS}{non-line-of-sight}
\acro{NR}{new radio}
\acro{OFDM}{orthogonal frequency-division multiplexing}
\acro{p.d.f.}{probability distribution function}
\acro{PAM}{pulse amplitude modulation}
\acro{PSD}{power spectral density}
\acro{PSK}{phase shift keying}
\acro{QAM}{quadrature amplitude modulation}
\acro{QPSK}{quadrature phase shift keying}
\acro{RCS}{radar cross-section}
\acro{RE}{resource element}
\acro{IRS}{intelligent reflecting surface}
\acro{r.v.}{random variable}
\acro{R.V.}{random vector}
\acro{SINR}{signal-to-interference plus noise ratio}
\acro{EJR}{echo-to-jammer ratio}
\acro{SJR}{signal-to-jammer ratio}
\acro{SNR}{signal-to-noise ratio}
\acro{SS}{spread spectrum}
\acro{SSIR}{signal-to-self interference ratio}
\acro{SVM}{support vector machine}
\acro{TH}{time-hopping}
\acro{ToA}{time-of-arrival}
\acro{UE}{user equipment}
\acro{ULA}{uniform linear array}
\acro{UWB}{ultrawide band}
\acro{UWBcap}[UWB]{Ultrawide band}
\acro{VAE}{variational autoencoder}
\acro{VI}{variational inference}
\acro{WLAN}{wireless local area network}
\acro{WMAN}{wireless metropolitan area network}
\acro{WPAN}{wireless personal area network}
\acro{WSN}{wireless sensor network}

\acro{ML}{machine learning}
\acro{GLR}{generalized likelihood ratio}
\acro{GLRT}{generalized likelihood ratio test}
\acro{LLRT}{log-likelihood ratio test}
\acro{LRT}{likelihood ratio test}
\acro{$P_{EM}$}{probability of emulation, or false alarm}
\acro{$P_{MD}$}{probability of missed detection}
\acro{$P_D$}{probability of detection}
\acro{$P_{FA}$}{probability of false alarm}
\acro{ROC}{receiver operating characteristic}

\acro{5G}{5th generation}
\acro{analog-digital converter}{ADC}
\acro{AGM}{arithmetic-geometric mean ratio test}
\acro{AI}{artificial intelligence}
\acro{AIC}{Akaike information criterion}
\acro{BMA}{basic mass assignment}
\acro{BIC}{Bayesian information criterion}
\acro{BWB}{bounded worse beaviour}
\acro{CA-CFAR}{cell-averaging constant false alarm rate}
\acro{CAIC}{consistent AIC}
\acro{CAICF}{consistent AIC with Fisher information}
\acro{CAVI}{coordinate ascent variational inference}
\acro{CFAR}{constant false alarm rate}
\acro{CDR}{constant detection rate}
\acro{CP}{cyclic prefix}
\acro{CPC}{cognitive pilot channel}
\acro{CL}{centroid localization}
\acro{CNN}{convolutional neural network}
\acro{CSCG}{circularly-symmetric complex Gaussian}
\acro{d.o.f.}{degrees of freedom}
\acro{DC}{distance based combining}
\acro{DNN}{deep neural network}
\acro{DSM}{dynamic spectrum management}
\acro{DSA}{dynamic spectrum access}
\acro{DTV}{digital television}
\acro{ECC}{European Communications Committee}
\acro{ED}{energy detector}
\acro{EDC}{efficient detection criterion}
\acro{EGC}{equal gain combining}
\acro{ELBO}{evidence lower bound}
\acro{EME}{energy to minimum eigenvalue ratio}
\acro{ENP}{estimated noise power}
\acro{ERC}{European Research Council}
\acro{ES-WCL}{energy based selection WCL}
\acro{EU}{European Union}
\acro{ExAIC}{exact AIC}
\acro{ExBIC}{exact BIC}
\acro{FC}{fusion center}
\acro{FIM}{Fisher information matrix}
\acro{FL}{federated learning}
\acro{FPGA}{field-programmable gate array}
\acro{GLRT}{generalized likelihood ratio test}
\acro{GIC}{generalized information criterion}
\acro{GMM}{Gaussian mixture model}
\acro{HC}{hard combining}
\acro{HD}{hard decision}
\acro{HF}{hard fusion}
\acro{tanh}{hyperbolic tangent}
\acro{i.i.d.}{independent, identically distributed}
\acro{ISM}{industrial, scientific, and medical}
\acro{ITC}{information theoretic criteria}
\acro{K-L}{Kullback-Leibler}
\acro{LR}{likelihood ratio}
\acro{LS}{least square}
\acro{LTE}{long term evolution}
\acro{MDL}{minimum description length}
\acro{MET}{ratio of maximum eigenvalue to the trace}
\acro{MOS}{model order selection}
\acro{MT}{multitaper}
\acro{MME}{maximum to minimum eigenvalues ratio}
\acro{MSE}{mean squared error}
\acro{mmWave}{millimeter-wave}
\acro{NP}{Neyman-Pearson}
\acro{NN}{neural network}
\acro{NRMSE}{normalized root mean square error}
\acro{p.d.f.}{probability density function}
\acro{OFCOM}{Office of Communications}
\acro{OSA}{opportunistic spectrum access}
\acro{OSF}{oversampling factor}
\acro{PCA}{principal component analysis}
\acro{PMR}{professional mobile radio}
\acro{PMSE}{programme making and special events}
\acro{PU}{primary user}
\acro{PWCL}{power based WCL}
\acro{r.o.f.}{roll-off factor}
\acro{RAN}{Regional Area Network}
\acro{ReLU}{rectified linear unit}
\acro{RF}{radio frequency}
\acro{RMSE}{root mean square error}
\acro{RWCL}{relative WCL}
\acro{RSS}{received signal strength}
\acro{SC}{soft combining}
\acro{SCM}{sample covariance matrix}
\acro{SCovM}{sample covariance matrix}
\acro{SCorM}{sample correlation matrix}
\acro{SD}{soft decision}
\acro{SDR}{software defined radio}
\acro{SF}{soft fusion}
\acro{SP}{sensing period}
\acro{SS}{spectrum sensing}
\acro{SSE}{signal subspace eigenvalues}
\acro{SU}{secondary user}
\acro{TOA}{time of arrival}
\acro{TDOA}{time difference of arrival}
\acro{TVBD}{TV Bands Device}
\acro{TS-WCL}{two step WCL}
\acro{VBFA}{variational Bayes factor analysis}
\acro{WCL}{weighted centroid localization}
\acro{WOSA}{weighted overlapped segment averaging}
\acro{WSS}{wideband spectrum sensing}
\acro{WSD}{white space device}
\acro{WSN}{wireless sensor network}
\acro{WSS}{wideband spectrum sensing}

\end{acronym}
  
\IEEEoverridecommandlockouts

\begin{document}
\title{Jamming Detection in MIMO-OFDM ISAC Systems Using Variational Autoencoders}
\author{
    \IEEEauthorblockN{Luca~Arcangeloni, Enrico~Testi, and Andrea~Giorgetti}
    \IEEEauthorblockA{CNIT/WiLab, DEI, University of Bologna, Italy
    \\Email:\{luca.arcangeloni2, enrico.testi, andrea.giorgetti\}@unibo.it}
\thanks{This work was supported by the European Union under the Italian National Recovery and Resilience Plan (NRRP) of NextGenerationEU, partnership on ``Telecommunications of the Future'' (PE00000001 - program ``RESTART'').
The authors are with the Department of Electrical, Electronic, and Information Engineering ``Guglielmo Marconi'' (DEI), CNIT, University of Bologna, Italy (e-mail: \{luca.arcangeloni2, enrico.testi, andrea.giorgetti\}@unibo.it).}%
}
\maketitle
\begin{abstract}
This paper introduces a novel unsupervised jamming detection framework designed specifically for monostatic \ac{MIMO}-\ac{OFDM} radar systems. The framework leverages echo signals captured at the \ac{BS} and employs the latent data representation learning capability of \acp{VAE}. The \ac{VAE}-based detector is trained on echo signals received from a real target in the absence of jamming, enabling it to learn an optimal latent representation of normal network operation. During testing, in the presence of a jammer, the detector identifies anomalous signals by their inability to conform to the learned latent space. We assess the performance of the proposed method in a typical \ac{ISAC}-enabled 5G wireless network, even comparing it with a conventional autoencoder.
\acresetall
\end{abstract}
%

\section{Introduction}\label{sec.intro}
\IEEEPARstart{I}{n} the context of 6G, the emerging paradigm of \ac{ISAC} systems is expected to revolutionize applications such as traffic monitoring and autonomous driving, while also enhancing urban safety \cite{WeiLuiMas:M22,DarMizSil:J24,SilDarMiz:C24}. Despite these advantages, \ac{ISAC} systems are sensitive to a range of traditional radar security threats (e.g., advanced radar \ac{ECM}), which are designed to deceive the sensing systems \cite{ZhaWanJia:J23}.
Among those attacks, the deceptive jamming technique known as \ac{DRFM} is particularly significant, as it enables precise scaling and delaying of intercepted radar waveforms by the jammer. Moreover, the global \ac{DRFM} market is projected to experience substantial growth due to the widespread adoption of \ac{AI}.
A deceptive jammer can exploit information about signals transmitted by the \ac{BS} to mimic the behavior of a typical network \ac{UE}. For example, common pilot/reference signals proposed for integrating sensing capabilities into communication networks \cite{ZhaRahWu:J22, WilBraVis:21,HuaWanLiu:J22} might already be known to intruders, potentially jeopardizing network security.
%

In radar literature, several approaches have been proposed to detect a deceptive jammer. Classic methods exploit likelihood-based algorithms that model the echo signals and adopt the \ac{GLRT} \cite{MarFulAlf:J08,ZhaZhoZha:J17}. They employ a two-block approach to unravel a multiple hypothesis test: initially, the presence of a target is detected; then, a second test is conducted to distinguish between an actual radar target and a false one. While such likelihood-based methods rely on prior information such as the statistical distribution of the channel and that of the clutter, in \cite{ChaLeiYin:C19} jamming detection and classification are performed by means of a decision tree and a \ac{SVM}, fed with a set of features extracted directly from the received signal.
In \cite{WanZhaWan:J20}, the authors propose \ac{ECCM} schemes for \ac{OFDM} radar that improve local \ac{SINR}, optimize initial phases to resist deception jamming, and develop waveform optimization methods to minimize jamming energy.
In \cite{SunYuaFul:C24}, the authors propose a power optimization strategy for multiple radar systems to counteract deception jamming in multi-target tracking tasks. They derive the posterior Cramér-Rao lower bound for deception range, which is crucial for distinguishing between physical and false targets. Using this metric, they introduce a method for false target discrimination and formulate a power optimization problem aimed at optimizing both tracking accuracy and discrimination performance.
Recently, the introduction of \ac{AI} techniques in the field of wireless communications gave impetus to the development of \ac{NN}-based jamming detectors \cite{ArcTesGio:J23,TesGio:M24}. In \cite{LvQuaFen:J21}, an ensemble \ac{CNN} with transfer learning is proposed to recognize a variety of deceptive jammers. In particular, a time-frequency dataset is constructed using the short-time Fourier transform. Then, features such as the real and imaginary parts, modulus, and phase are extracted to build different subdatasets. These subdatasets serve as the input for classifiers in the ensemble model.

In this work, we present a novel unsupervised jamming detection framework tailored for a monostatic \ac{MIMO}-\ac{OFDM} radar system.
The framework utilizes the echo signals captured at the \ac{BS} and employs the \ac{VAE}, a generative latent variable model adept at learning latent data representations \cite{KinWell:J13}. The contributions of this work are the following:
\begin{itemize} 
    \item We propose a jamming detection framework that, starting from the echo signals acquired by the \ac{BS} and leveraging the latent space identification capability of \acp{VAE} is able to detect the presence of an intruder in a \ac{MIMO}-\ac{OFDM} system. The detector is trained on a dataset consisting of received echoes from a real target in the absence of jamming, allowing the \ac{VAE} to learn the optimal latent representation of the data. During testing, when the jammer is present, the detector identifies the anomalous signal by its incompatibility with the learned latent space.

    \item To evaluate the performance of the proposed solution, we investigate a case study consisting of a 5G wireless network employing an \ac{ISAC} system in presence of a deceptive jammer.

    \item Finally, the performance of the \ac{VAE}-based jamming detector is compared with that of a conventional \ac{AE}, which has been carefully designed and trained to achieve its optimal performance.
\end{itemize} 

Throughout the paper, capital and lowercase boldface letters denote matrices and vectors, respectively. With $v_{i,j}$, $\mathbf{v}_{i,:}$, and $\mathbf{v}_{:,j}$, we represent, respectively, the element, the $i$th row, and the $j$th column of the matrix $\mathbf{V}$.
$X\thicksim \mathcal{U}\!\left(a,b\right)$ denotes a uniform distributed \ac{r.v.} between $a$ and $b$. $X \thicksim \mathcal{N}(\mu,\sigma^2)$ denotes a Gaussian \ac{r.v.} with mean $\mu$ and variance $\sigma^2$, and $Z \thicksim \mathcal{CN}(0,\sigma^2)$  denotes a zero-mean circularly symmetric complex Gaussian \ac{r.v.} with variance $\sigma^2$. 
We denote the expectation operator by $\mathbb{E}[X]$.
$\trans{(\cdot)}$ and $(\cdot)^*$ denote the transpose and the conjugate operations, respectively.

\section{System Model}\label{sec:sysmodel}

\begin{figure*}[t]
	\centering
	\includegraphics[width=0.99\textwidth,draft=false]{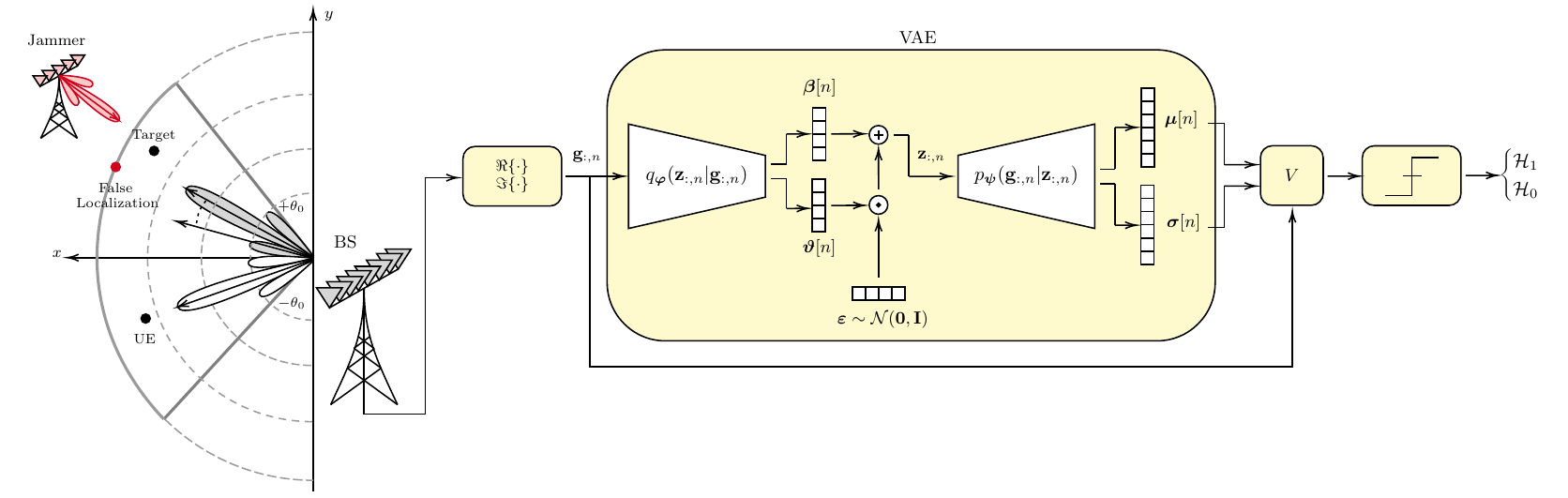}
	\caption{The monostatic \ac{OFDM} \ac{ISAC} scheme in presence of a jammer: an illustration of the data gathering process and decision making. The latent variable inside the \ac{VAE} is obtained using the well-known reparameterization trick $\mathbf{z}_{:,n}=\bm{\beta}[n] + \bm{\vartheta}[n]\cdot\bm{\varepsilon}$, where $\mathbf{c}\cdot \mathbf{d}$ is the element-wise product between the two vectors $\mathbf{c}$ and $\mathbf{d}$.}
	\label{fig:scenario}
\end{figure*}

Let us consider the monostatic \ac{MIMO}-\ac{OFDM} system depicted in Fig.~\ref{fig:scenario}, which consists of transmitter and receiver antenna arrays with $\Nt$ and $\Nr$ elements, respectively, used for communication and sensing. We assume that \acp{ULA} with half-wavelength separation, i.e., $d = \lambda/2$, where $\lambda = c/f_\mathrm{c}$, $c$ is the speed of light, and $f_\mathrm{c}$ is the carrier frequency, are employed for both transmission and reception. According to \cite{EliEliPuc:J23}, we assume that sensing is performed using repeated time-frequency slots composed of $K$ subcarriers and $M$ \ac{OFDM} symbols each. Within such slots, a sensing beam is activated beside a communication beam (for downlink communication towards a user) \cite{PucPaoGio:J22}. However, for jamming detection, we pick one ``observation'' within each slot, which refers to a vector containing the received OFDM symbols (right after the FFT processing at the sensing receiver) across the $K$ subcarriers at time $n$.

With such a multibeam \ac{ISAC} approach only a fraction of total power of the \ac{OFDM} signal is designated to sensing purposes.
The discrete-time transmitted signal in the $k$th subcarrier of at time $n$, can be written as \cite{PucPaoGio:J22}
\begin{equation}
    \x[k,n] = \mathbf{w}_{\text{T}}[n]x_{k,n}
\end{equation}
where $k=1,\dots,K$, $n=1,\dots,N$ with $N$ the number of observations, and $\mathbf{w}_{\text{T}}[n] \in \mathbb{C}^{\Nt\times1}$ is the sensing beamforming vector used to map each modulation symbol, $x_{k,n}$, to the transmitting antennas.
By considering a beam steering approach and performing a normalization with respect to the \ac{EIRP} $P_{\text{T}}G_{\text{T}}$, we express the beamforming vector as
\begin{equation}
    \mathbf{w}_{\text{T}}[n] = \frac{\sqrt{\rho P_{\text{T}}G_{\text{T}}}}{\Nt} \mathbf{a}_{\text{T}}^*(\theta^n_{\text{T}})
\end{equation}
where $\rho \in [0, 1]$ is the parameter used to control the fraction of the total power apportioned to the sensing direction, $P_{\text{T}}$ is the transmit power, $G_{\text{T}}$ is the transmit array gain along the beam steering direction, and $\mathbf{a}_{\text{T}}(\theta^n_{\text{T}}) \in \mathbb{C}^{\Nt\times1}$ is the steering vector along the sensing directions $\theta^n_{\text{T}}$.  In particular, the steering vector for the considered \ac{ULA} can be expressed as
\begin{equation}
    \mathbf{a}_{\text{T}}(\theta^n_{\text{T}}) = \left[ 1, e^{i\pi\sin(\theta^n_{\text{T}})},\dots,e^{i\pi(\Nt-1)\sin(\theta^n_{\text{T}})}\right]^{\mathsf{T}}.
\end{equation}

For generality and to facilitate jamming detection in an unknown environment, we assume the interval between two consecutive observations exceeds the channel's coherence time. This results in different channel realizations for the sensing receiver with each observation. Additionally, the characteristics of both the target and jammer may vary across observations; for instance, their positions relative to the \ac{BS} may change.


\subsection{Jammer Model}\label{subsec:jammer}
A deceptive jammer in radar systems is a sophisticated \ac{ECM} designed to mislead radar operations by transmitting false signals. These jammers create fake targets or alter the perceived location and speed of actual targets, confusing the radar's tracking and detection capabilities.
The most efficient way to implement a deceptive jammer consists of using a \ac{DRFM}. It is a high-speed, analog-to-digital converter and storage system that provides the capability to sample, process, and playback \ac{RF} signals with minimum loss of fidelity.
With these capabilities, the jammer can rapidly obtain pertinent data and formulate an effective strategy. For instance, in a hypothetical 5G NR network where \ac{ISAC} is performed using the \ac{PDSCH-DMRS}, also designated as the pilot, the jammer may be able to access the primary and secondary synchronization signals in order to obtain the physical cell identity. The cell ID comprises data regarding the initialization of the \ac{PDSCH-DMRS}. Consequently, the jammer is aware of the \acp{RE} where the \ac{PDSCH-DMRS} is transmitted and what symbols it is composed of, thus facilitating the transmission of a delayed version of the pilot towards the \ac{BS}.

Let us consider a deceptive jammer capable of mimicking the signal transmitted by the \ac{BS} and injecting a false delay into the received signal with the aim of falsifying the \ac{BS}'s estimated location of the target. it can quickly obtain this information and plan its response accordingly. Specifically, this study assumes that the jamming attack occurs in the $n$th observation and $k$th subcarrier, using the same symbols $x_{k,n}$ transmitted by the \ac{BS}. This represents a worst-case scenario where the radar receiver is completely misled.\footnote{For example, if sensing is performed using pilot symbols the possibility that the jammer might know $x_{k,n}$ is not unlikely.}

The jammer comprises a transmitter antenna array with $\Nj$ elements arranged in an \ac{ULA} with half-wavelength separation and adopts \ac{OFDM} modulation with $K$ subcarriers. Therefore, its signal can be written as 
\begin{align}
    \Tilde{\mathbf{x}}[k,n] = \mathbf{w}_{\text{J}}[n]x_{k,n}e^{-i2\pi k\Delta f\tau_\mathrm{f}^n}
\end{align}
where $\Delta f$ is the subcarrier spacing, $\mathbf{w}_{\text{J}}[n]\in\mathbb{C}^{\Nj\times1}$ is the jammer beamforming vector, and $\tau_\mathrm{f}^n$ is the false delay introduced by the jammer.
The beamforming vector can be expressed as
\begin{equation}
    \mathbf{w}_{\text{J}}[n] = \frac{\sqrt{P_{\text{J}}G_{\text{J}}}}{\Nj}\mathbf{a}^*_{\text{J}}(\theta_{\text{J}}^n)
\end{equation}
where $P_{\text{J}}$ is the jammer signal power, $G_{\text{J}}$ is the array gain along the beam steering direction, and $\mathbf{a}_{\text{J}}(\theta^n_{\text{J}}) \in \mathbb{C}^{\Nj\times1}$ is the steering vector.

\subsection{Received Signal at the BS}
Let us assume the presence of a point-like target within the sensing beam. The received signal is processed by a typical \ac{OFDM} receiver \cite{PucPaoGio:J22}, such that the vector $\y[k,n]\in \mathbb{C}^{\Nr\times1}$ of the received modulation symbols at each antenna after the \ac{FFT} block is 
\begin{align} \label{eq:YwithJ}
    \y[k,n] &= \mathbf{H}[k,n]\x[k,n] + \Tilde{\mathbf{H}}[k,n] \Tilde{\mathbf{x}}[k,n] \nonumber \\
    &+ \bm{\nu}[k,n] + \mathbf{m}[k,n]
\end{align}
where $\mathbf{H}[k,n] \in\mathbb{C}^{\Nr\times\Nt}$ is the channel matrix  between the target and the \ac{BS} for the $k$th subcarrier in the $n$th observation, $\Tilde{\mathbf{H}}[k,n] \in\mathbb{C}^{\Nr\times\Nj}$ is the channel matrix between the \ac{BS} and the jammer, $\bm{\nu}[k,n]\in\mathbb{C}^{\Nr\times1}$ is the vector whose elements represent the self-interference due to imperfect Tx–Rx isolation at each receiving antenna, and $\mathbf{m}[k,n] \thicksim \mathcal{CN}(\mathbf{0},\sigma^2_{\text{N}}\mathbf{I}_{\Nr})$ is the noise power at the sensing receiver. Let us remark that, in absence of the jammer, the second term in \eqref{eq:YwithJ} is zero.

The \ac{BS}-target-\ac{BS} channel matrix can be written as \cite{PucPaoGio:J22}\footnote{In this work we do not consider the Doppler effect for brevity but its presence can be included in the system and jammer models without altering the operating principle of the jammer detector.}
\begin{align} \label{eq:Htmp}
    \mathbf{H}[k,n] = \alpha_\mathrm{t}^n e^{i\phi_\mathrm{t}^n}e^{-i2\pi k \Delta f \tau_\mathrm{t}^n}\mathbf{a}_\mathrm{R}(\theta_\mathrm{t}^n) \mathbf{a}_\mathrm{T}^{\mathsf{T}}(\theta_\mathrm{t}^n)
\end{align}
where $\alpha_\mathrm{t}^n$, $\phi_\mathrm{t}^n$, $\tau_\mathrm{t}^n$, $\theta_\mathrm{t}^n$ are the attenuation, phase, delay, and \ac{AoA}/\ac{AoD} of the target for the $n$th observation, respectively.

The gain $\alpha_\mathrm{t}^n$ includes the attenuation along the \ac{BS}-target-\ac{BS} path, that is calculated as
\begin{equation}
    \alpha_\mathrm{t}^n = \sqrt{\frac{c^2\sigma^n_{\text{RCS}}}{(4\pi)^3f_\mathrm{c}^2(r_\mathrm{t}^n)^4}}
\end{equation}
where $r_\mathrm{t}^n$ is the distance between the target and the \ac{BS} in the $n$th observation, and $\sigma^n_{\text{RCS}}$ is its \ac{RCS}. The target \ac{RCS} is assumed to adhere to the Swerling~I model, i.e., $\sigma^n_{\text{RCS}} \thicksim \text{exp}(\Bar{\sigma}_{\text{RCS}})$.

The jammer-\ac{BS} channel matrix, instead, can be written as
\begin{align}
    \Tilde{\mathbf{H}}[k,n] = \alpha_\mathrm{J}^n e^{i\phi_\mathrm{J}^n} e^{-i2\pi k\Delta f\tau_\mathrm{J}^n}\mathbf{a}_\mathrm{R}(\theta_\mathrm{BS,J}^n) \mathbf{a}_\text{J}^{\mathsf{T}}(\theta_\mathrm{J,BS}^n)
\end{align}
where $\alpha_\mathrm{J}^n$ is the channel attenuation given by the path-loss equation, $\phi_\mathrm{J}^n$ is the phase shift, and $\tau_\mathrm{J}^n$ is the delay of the direct path. Furthermore, $\theta_\mathrm{BS,J}^n$ and $\theta_\mathrm{J,BS}^n$ are the \ac{AoA} and \ac{AoD} of the \ac{LOS} path.

Regarding the self-interference term in \eqref{eq:YwithJ}, each element of vector $\bm{\nu}[k,n]$ can be interpreted as the signal scattered by a static target located very close to the receiver \cite{PucPaoGio:J22}. Hence, 
\begin{align}
    \bm{\nu}[k,n] = \alpha^n_{\text{SI}}x_{k,n}[e^{i\phi^n_{\text{SI},1}},\dots,e^{i\phi^n_{\text{SI},\Nr}}]^{\mathsf{T}}
\end{align}
where $\alpha^n_{\text{SI}}$ is the self-interference attenuation and is the same for all receiving antennas, and $[\phi^n_{\text{SI},1},\dots,\phi^n_{\text{SI},\Nr}]$ are the phase shifts at the antennas. 

\subsection{Processing at the Base Station}
Let us consider the vector of received symbols obtained from \eqref{eq:YwithJ}, $\y[k,n]$, and let us assume a specific sensing direction such that $\theta^n_{\text{R}} = \theta^n_{\text{T}}$. Spatial combining is then performed using the receiving beamforming vector as
\begin{equation*}
    \mathbf{w}_{\text{R}}[n]= \mathbf{a}_{\text{R}}^*(\theta^n_{\text{R}}) = \left[ 1, e^{-i\pi\sin(\theta^n_{\text{R}})},\dots,e^{-i\pi(\Nr-1)\sin(\theta^n_{\text{R}})}\right]^{\mathsf{T}}.
\end{equation*}
This results in the formation of a grid of received symbols, where each element $y_{k,n}$ is obtained by taking the inner product between the receiving beamforming vector $\mathbf{w}_{\text{R}}[n]$ and the vector of the symbols received at each antenna $\y[k,n]$, i.e., $y_{k,n} = \mathbf{w}_{\text{R}}^{\mathsf{T}}[n]\y[k,n]$.
Then, reciprocal filtering is performed, which consists of an element-wise division between the received and the transmitted grids to remove the dependence on the transmitted symbols, yielding $g_{k,n} = y_{k,n} / x_{k,n}$. Considering the $n$th observation, we have 
\begin{align}
    g_{k,n} &= \mathbf{w}_{\text{R}}^{\mathsf{T}}[n]\mathbf{H}[k,n] \mathbf{w}_{\text{T}}[n] \nonumber \\ 
    &+\, \mathbf{w}_{\text{R}}^{\mathsf{T}}[n]\Tilde{\mathbf{H}}[k,n]\mathbf{w}_{\text{J}}[n] e^{-i2\pi k\Delta f\tau_\mathrm{f}^n} \nonumber \\ 
    &+ \frac{\mathbf{w}_{\text{R}}^{\mathsf{T}}[n]\bm{\nu}[k,n]}{x_{k,n}} + \frac{\mathbf{w}_{\text{R}}^{\mathsf{T}}[n]\mathbf{m}[k,n]}{x_{k,n}} \label{eq:gJ}
\end{align}
where the second term is the signal injected by the jammer into the \ac{BS} receiver to deceive sensing. 
Finally, for the sake of jamming detection, the real and imaginary parts of $g_{k,n}$ are split and arranged in a matrix $\mathbf{G} \in \mathbb{C}^{2K\times N}$, whose $n$th column is
\begin{equation*}
    \mathbf{g}_{:,n} = \left[\Re\{g_{0,n}\},\dots,\Re\{g_{K,n}\},\Im\{g_{0,n}\},\dots,\Im\{g_{K,n}\}\right]^{\mathsf{T}}.
\end{equation*}
\section{Variational Autoencoder} \label{sec:vae}
In this section, we introduce a variational Bayesian approach to detect the presence of the deceptive jammer introduced in Section~\ref{subsec:jammer}.
We begin our analysis by observing that the system model in \eqref{eq:gJ}, in the absence of a jammer (i.e., when the second term is zero), can be interpreted as a latent model for the generation of $g_{k,n}$. 
Our initial objective is to learn the latent space generated by the system under no-jammer conditions by means of a variational autoencoder. Subsequently, we perform jamming detection based on the premise that the presence of a jammer will significantly alter the latent space, rendering it markedly different from the one identified in a jammer-free environment.

\subsection{Variational autoencoder}
\Ac{VAE} is a popular approach for unsupervised learning of generative latent models exploiting the power of neural networks \cite{KinWell:J13}. In other words, it is a modern formulation of the \ac{VI} framework, where the goal is to provide a good approximation for the posterior distribution $p_{\bm{\psi}}(\mathbf{Z}|\mathbf{G})$ of the latent variables $\mathbf{Z}=\{\mathbf{z}_{:,n}\}_{n=1}^N$ given the observed data $\mathbf{G}$ and with parameters $\bm{\psi}$ \cite{Hoff:J13}. Assuming that the observations in $\mathbf{G}$ are independent, we aim to find $p_{\bm{\psi}}(\mathbf{z}_{:,n}|\mathbf{g}_{:,n})$.
Since the posterior cannot be directly computed due to the intractability of the marginal likelihood, \ac{VI} provides an approximate distribution $q_{\bm{\varphi}}(\mathbf{z}_{:,n}|\mathbf{g}_{:,n})$ with parameters $\bm{\varphi}$. In particular, the best approximation can be computed by minimizing the following Kullback-Leibler divergence
\begin{align} \label{eq:lb}
    D_{\text{KL}}& (q_{\bm{\varphi}}(\mathbf{z}_{:,n}|\mathbf{g}_{:,n})||p_{\bm{\psi}}(\mathbf{z}_{:,n}|\mathbf{g}_{:,n})) = \nonumber \\
    &- \mathbb{E}_{q_{\bm{\varphi}}(\mathbf{z}_{:,n}|\mathbf{g}_{:,n})}\left[\ln\frac{p_{\bm{\psi}}(\mathbf{g}_{:,n},\mathbf{z}_{:,n})}{q_{\bm{\varphi}}(\mathbf{z}_{:,n}|\mathbf{g}_{:,n})}\right] + \ln p_{\bm{\psi}}(\mathbf{g}_{:,n}) 
\end{align}
or, equally, by maximizing the \ac{ELBO}
\begin{align}
    \mathcal{L}(\bm{\psi},\bm{\varphi},\mathbf{g}_{:,n}) = \mathbb{E}_{q_{\bm{\varphi}}(\mathbf{z}_{:,n}|\mathbf{g}_{:,n})}\left[\ln\frac{p_{\bm{\psi}}(\mathbf{g}_{:,n},\mathbf{z}_{:,n})}{q_{\bm{\varphi}}(\mathbf{z}_{:,n}|\mathbf{g}_{:,n})}\right].
\end{align}

\Acp{VAE} provides a stochastic \ac{VI} solution aiming to maximize the $\mathcal{L}(\bm{\psi},\bm{\varphi},\mathbf{g}_{:,n})$ by means of gradient-based optimization.

Let us now assume the following prior distributions
\begin{align}
    p_{\bm{\psi}}(\mathbf{g}_{:,n}|\mathbf{z}_{:,n}) &\thicksim \mathcal{N}(\bm{\mu}[n],\bm{\sigma}^2[n]\mathbf{I}_{2K}), \label{eq:sz} \\
    p(\mathbf{z}_{:,n}) &\thicksim \mathcal{N}(\mathbf{0},\mathbf{I}_{L}), \label{eq:z} \\
    q_{\bm{\varphi}}(\mathbf{z}_{:,n}|\mathbf{g}_{:,n}) &\thicksim \mathcal{N}(\bm{\beta}[n],\bm{\vartheta}^2[n]\mathbf{I}_{L}). \label{eq:zs}
\end{align}
The choice of Gaussian priors for the latent space has many benefits: (i) the Gaussian distribution is mathematically convenient due to its properties, such as admitting a closed-form expression for the Kullback-Leibler divergence, which is essential for the \ac{VAE}'s optimization process; (ii) thanks to the central limit theorem, Gaussian priors are a natural and generalizable choice for modeling the latent space of diverse datasets; (iii) empirically, Gaussian priors have been shown to produce smooth and continuous latent spaces, which are desirable for generative tasks \cite{KinWell:J13}.
After proper manipulations, the \ac{ELBO} can be written as
\begin{align}
    \mathcal{L}(\bm{\psi},\bm{\varphi},\mathbf{g}_{:,n}) &= -D_{\text{KL}}\left(q_{\bm{\varphi}}(\mathbf{z}_{:,n}|\mathbf{g}_{:,n})||p(\mathbf{z}_{:,n})\right) \nonumber \\
    & + \mathbb{E}_{q_{\bm{\varphi}}(\mathbf{z}_{:,n}|\mathbf{g}_{:,n})}\left[\ln p_{\bm{\psi}}(\mathbf{g}_{:,n}|\mathbf{z}_{:,n})\right]. \label{eq:elbofirst}
\end{align}
Adopting the reparameterization trick proposed in \cite{KinWell:J13}, considering the prior distributions in \eqref{eq:sz}, \eqref{eq:z}, and \eqref{eq:zs}, it is possible to obtain a differentiable formulation for the \ac{ELBO}, i.e., 
\begin{align} \label{eq:elboFinal}
    \mathcal{L}&(\bm{\psi},\bm{\varphi},\mathbf{g}_{:,n}) = \frac{1}{2}\sum_{l=1}^L \left(1+\ln \vartheta_l^2[n] - \beta_l^2[n] - \vartheta_l^2[n] \right) \nonumber \\
    &-\underbrace{\frac{1}{2}\sum_{k=1}^{2K}\left(\ln2\pi + \ln\sigma_k^2[n] +\frac{(g_{k,n}-\mu_k[n])^2}{\sigma_k^2[n]} \right)}_\text{$V$}
\end{align}
where $V=-\mathbb{E}_{q_{\bm{\varphi}}(\mathbf{z}_{:,n}|\mathbf{g}_{:,n})}\left[\ln p_{\bm{\psi}}(\mathbf{g}_{:,n}|\mathbf{z}_{:,n})\right]$ is the reconstruction probability that will be used for jamming detection \cite{AnCho:J15}.

Fig.~\ref{fig:scenario} encloses a schematic representation of a \ac{VAE}. The function $q_{\bm{\varphi}}(\mathbf{z}_{:,n}|\mathbf{g}_{:,n})$ serves as a probabilistic encoder that, given an input $\mathbf{g}_{:,n}$, generates a distribution over the possible values of $\mathbf{z}_{:,n}$ from which $\mathbf{g}_{:,n}$ could have been produced. Similarly, $p_{\bm{\psi}}(\mathbf{g}_{:,n}|\mathbf{z}_{:,n})$ functions as a probabilistic decoder, producing a distribution over the possible values of $\mathbf{g}_{:,n}$ given $\mathbf{z}_{:,n}$. Thus, $\bm{\mu}[n]$, $\bm{\sigma}[n]$, $\bm{\beta}[n]$, and $\bm{\vartheta}[n]$ are the outputs of the encoder and decoder neural networks, whose weights are denoted by $\bm{\psi}$ and $\bm{\varphi}$, respectively.

\subsection{Anomaly detection} \label{sec:ad}
The \ac{VAE}, trained on echoes captured in absence of a jammer, seeks to learn the latent variables that better represent the channel in presence of a target. Thus, after the training, if the \ac{VAE} is fed with a vector $\mathbf{g}_{:,n}$ corresponding to the manipulated received signal in presence of a jammer, it provides an anomalous value for $\mathcal{L}(\bm{\psi},\bm{\varphi},\mathbf{g}_{:,n})$. 
Specifically, when the jammer is present results in significantly high values of the reconstruction probability $V$.
Therefore, based on such considerations, we propose an anomaly detector that employs as metric the reconstruction probability, i.e.,
\begin{equation} \label{eq:test2} 
    V \stackrel[\mathcal{H}_0]{\mathcal{H}_1}{\gtrless} \omega.
\end{equation}
Hypothesis $\mathcal{H}_1$ corresponds to the presence of a jammer, while the null hypothesis, $\mathcal{H}_0$, corresponds to its absence. The thresholds $\omega$ is obtained by setting the false alarm probability $p_{\text{FA}}=(V > \omega | \mathcal{H}_0)$, where the null distribution is calculated via histogram-based probability density function estimation.

\section{Numerical Results}\label{sec:numerical_results}
In this section, we evaluate the performance of the proposed solution and compare it with that of a conventional \ac{AE}. 
For all the simulations, 5G \ac{NR} signals compliant with 3GPP Technical Specification in \cite{5G} are considered. According to the 5G \ac{NR} standard, we employed a carrier frequency of $f_\mathrm{c}=28\,$GHz, an \ac{EIRP} $P_{\text{T}}G_{\text{T}}=13\,$dBW, subcarrier spacing $\Delta f=120\,$kHz, number of antennas $\Nt=\Nr=50$, and the number of subcarriers used for the radar set to $K=500$.
In addition, a \ac{QPSK} modulation alphabet is used for the generation of the \ac{OFDM} signal, and the parameter $\rho$ is set to $0.5$. As shown in Fig.~\ref{fig:scenario}, the system scans the environment in the range $[-\theta_0, \theta_0]$, with $\theta_0=60^{\circ}$ and a beamwidth $\Delta \Theta = 5.3^{\circ}$ at $-10\,$dB gain relative to the beam direction. Therefore, the number of step to cover the entire range is $N_\text{step}=\lceil \frac{2\theta_0}{\Delta\Theta}\rceil = 23$.

The self-interference attenuation $\alpha_{\text{SI}}$ is computed using the \ac{SSIR} defined as $\mathsf{SSIR} =(\alpha_\mathrm{t}^n/\alpha^n_{\text{SI}})^2=20\,$dB.
The mean \ac{RCS} is $\Bar{\sigma}_\text{RCS}=1\,$m$^2$, and the noise power spectral density is $N_0=k_\text{B}T_0F$, where $k_\text{B} = 1.38 \cdot 10^{-23}\,$JK$^{-1}$ is the Boltzmann constant, $T_0 = 290\,$K is the reference temperature, and $F_{\text{dB}} = 8\,$dB is the receiver noise figure. 

\subsection{Algorithm's parameters}\label{sec:net}
The \ac{VAE}'s encoder comprises a deep feed-forward neural network architecture, with the input layer receiving the normalized vector $\mathbf{g}_{:,n} \in \mathbb{R}^{2K\times1}$, which has unit modulus. Following this, the encoder employs $5$ hidden layers with  $728, 256, 64, 32,$ and $10$ neurons each, respectively. The encoder outputs two vectors, $\bm{\beta}[n]$ and $\bm{\vartheta}[n]$, each having a latent dimension of $L=10$.
The decoder takes the latent variable $\mathbf{z}_{:,n}=\bm{\beta}[n]+\bm{\vartheta}[n]\cdot\bm{\epsilon}$ as input, where $\bm{\epsilon} \thicksim \mathcal{N}(\bm{0},\mathbf{I}_\text{L})$. The decoder's architecture mirrors that of the encoder, producing the vectors $\bm{\mu}[n] \in\mathbb{R}^{2K\times1}$ and $\bm{\sigma}[n] \in\mathbb{R}^{2K\times1}$ as output.  
Each hidden layer employs the \ac{ReLU} activation function, except for the layers computing $\bm{\beta}[n]$ and $\bm{\vartheta}[n]$, which employ a linear activation function. Since the input is normalized with unit modulus, the output layer for $\bm{\mu}[n]$ adopts the hyperbolic tangent activation function.
Training is conducted using the Adagrad optimizer with learning rate $\eta=0.005$, for $N_\text{epoch}=4000$ epochs, and batch size $N_\text{bs}=460$. The training objective is to minimize the negative \ac{ELBO}, which is defined in \eqref{eq:elboFinal}.

To validate our \ac{VAE}-based approach, we compare its performance with a conventional \ac{AE}. The \ac{AE}'s encoder also employs a feed-forward deep neural network architecture, comprising $7$ hidden layers with $728, 512, 256, 128, 64, 32,$ and $10$ neurons each, respectively, where $10$ denotes the bottleneck dimension. The decoder mirrors the encoder's architecture, giving as output the reconstructed vector $\hat{\mathbf{g}}_{:,n}$. Each hidden layer employs the \ac{ReLU} activation function, while the output layer uses the hyperbolic tangent activation function.
The \ac{AE} is trained using the Adagrad optimizer with learning rate $\eta=0.001$, over $N_\text{epoch}=2000$ epochs, and a batch size of $N_\text{bs}=200$. The loss function used for the training is the \ac{MSE}. 

For both training and validation, we use a matrix $\mathbf{G}\in\mathbb{R}^{2K\times N}$ with $N=57.5\cdot10^3$ observations. Specifically, $80\%$ of the observations are used for training, and the remaining $20\%$ for validation. Both the \ac{VAE} and \ac{AE} architectures were selected after extensive parameter searches to achieve their optimal performance.
To obtain a comprehensive training set that enables the \ac{VAE} to learn a general representation of the latent space, we assume that independent observations are collected across various environments. Specifically, we assume that for each observation, the target position and the parameters related to the channel realization between the target and the \ac{BS} are generated according to the following distributions:
\begin{align}
    \phi_\mathrm{t}^n,\,\phi_{\text{SI},1},\,\dots,\,\phi_{\text{SI},\Nr} &\thicksim \mathcal{U}(0,2\pi), \label{eq:dist1} \\
    r_\mathrm{t}^n &\thicksim \mathcal{U}(20,85), \label{eq:dist2} \\
    \theta_\mathrm{t}^n,\,\theta_\mathrm{BS,J}^n &\thicksim \mathcal{U}[\theta^n_{\text{T}}-\Delta\Theta, \theta^n_{\text{T}}+\Delta\Theta]. \label{eq:dist3}
\end{align}
For the $n$th observation, the direction of the \ac{BS}' beam is set according to 
\begin{equation}
    \theta^n_{\text{T}}=\theta^n_{\text{R}}= -\theta_0 + \text{mod}(n-1, N_\text{step}) \Delta\Theta
\end{equation}
where $\text{mod}(a,b)$ is the modulo operator which returns the remainder of the division between the two positive numbers $a$ and $b$. 

To evaluate the efficacy of the anomaly detector, the input for the test is a matrix comprising $4600$ observations. Of these, $2300$ represent instances where only the target is present, while the remaining observations include both the target and the jammer. Also for the test dataset we assume that, for each observation, the target position and the parameters related to the channel realization between the target and the \ac{BS} are generated according to \eqref{eq:dist1}, \eqref{eq:dist2}, and \eqref{eq:dist3}, while the parameters related to the jammer are generated according to $\phi_\mathrm{J}^n \sim \mathcal{U}(0, 2\pi)$, $\theta_{\text{J}}^n \sim \mathcal{U}(0, 2\pi)$, and $\theta_\mathrm{J,BS}^n \sim \mathcal{U}[\theta_{\text{J}}^n - \Delta \Theta_j, \theta_{\text{J}}^n + \Delta \Theta_j]$, where $\Delta \Theta_j = 14^\circ$. The jammer is equipped with $\Nj=10$ antennas.

\begin{figure}[t]
	\centering
	\includegraphics[width=0.99\columnwidth,draft=false]{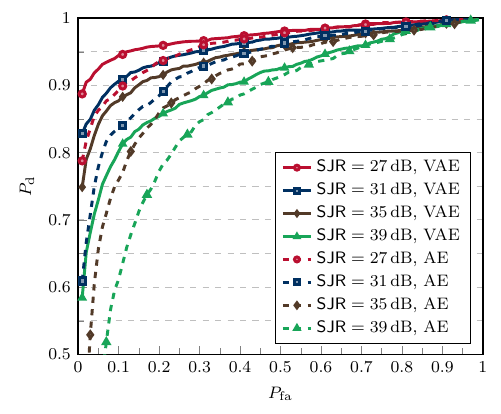}
	\caption{\Ac{ROC} curves of the proposed \ac{VAE} and the conventional \ac{AE} for different \ac{SJR} values.}
	\label{fig:rocsjr}
\end{figure}

\subsection{Impact of \ac{SJR}}\label{sec:sjr}
In this subsection, the performance of the proposed jamming detection method varying the \ac{SJR}, is studied. The \ac{SJR} is defined as the ratio between the \ac{EIRP} of the legitimate signal and that of the jammer signal, i.e.,
\begin{equation}
    \mathsf{SJR} = \frac{\rho P_{\text{T}}G_{\text{T}}}{P_{\text{J}}G_{\text{J}}}.
\end{equation}
The test is performed assuming the jammer is in a fixed position, i.e., $r_j^n=r_j=90\,$m, implying that the jammer is attempting to deceive the \ac{BS} by staying outside its coverage area.
For each observation, the injected false delay is set to $\tau_\mathrm{f}^n=0.17\,\mu$s, corresponding to a false distance of $50\,$m.
Fig.~\ref{fig:rocsjr} shows the \ac{ROC} curves for different $\mathsf{SJR}$ values for both the proposed \ac{VAE} and the conventional \ac{AE}. 
Considering a false alarm probability $P_\text{fa}=0.05$, the \ac{VAE} achieves a detection probability $P_\text{d}=0.93$ for $\mathsf{SJR}=27\,$dB. However, when the jammer's transmit power is significantly lower than \ac{BS}'s sensing power, the detection performance deteriorates. 
Moreover, from Fig.~\ref{fig:rocsjr} it is evident that the \ac{VAE} outperforms the conventional \ac{AE} for each of the $\mathsf{SJR}$ values. The best performance produced by the \ac{VAE} with regard to \ac{AE} are caused by the difference between reconstruction probability and reconstruction error. The latent variables in a \ac{VAE} are stochastic, whereas in autoencoders, they are defined by deterministic mappings. As the \ac{VAE} employs a probabilistic encoder to model the distribution of latent variables rather than the variables themselves, it is able to account for the variability of the latent space through the sampling process. This increases the expressive power of the \ac{VAE} in comparison to the autoencoder, as it is capable of capturing differences in variability even when normal and anomalous data share the same mean value \cite{AnCho:J15}.

\subsection{Latent space dimension}\label{sec:latdim}
\begin{figure}[t]
	\centering
	\includegraphics[width=0.99\columnwidth,draft=false]{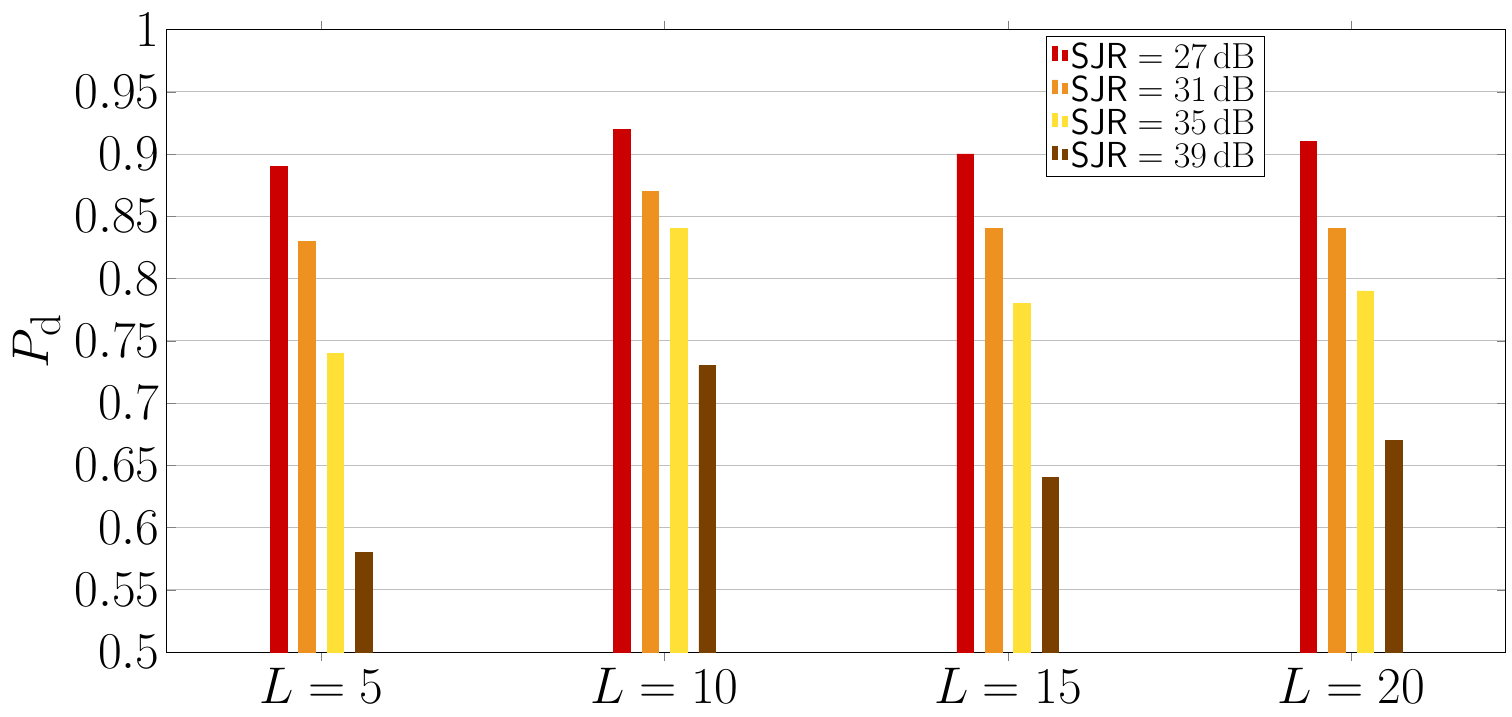}
	\caption{Probability of detection $P_\text{d}$ for different latent space dimensions, $L$, and \ac{SJR} values, with a false alarm probability $P_\text{fa}=0.05$.}
	\label{fig:histo}
\end{figure}

Finally, we assess the impact of the latent space dimension hyperparameter, $L$, on the \ac{VAE}'s detection performance. Fig.~\ref{fig:histo} shows the probability of detection $P_\text{d}$, for different $\mathsf{SJR}$ values and latent space dimensions $5$, $10$, $15$ and $20$, with a false alarm probability $P_\text{fa}=0.05$. Tipically, setting a low latent space dimension prevents the \ac{VAE} from capturing all the trends and variations in the training observations. Conversely, high values of $L$ tend to keep the regularization term $D_{\text{KL}}\left(q_{\bm{\varphi}}(\mathbf{z}_{:,n}|\mathbf{g}_{:,n})||p(\mathbf{z}_{:,n})\right)$ low during the training \cite{Car:T16}.
From Fig.~\ref{fig:histo}, it is evident that setting $L=10$ provides the best performance, even considering different \ac{SJR} values.
When $L=5$, the \ac{VAE} is unable to correctly learn the latent space, resulting in degraded detection probability. Similarly, for $L=15$ and $L=20$, the impact on the regularization term prevents the algorithm from fully exploiting its learning potential, leading to suboptimal performance. Therefore, both lower and higher values of $L$ negatively affect the detection probability.

\section{Conclusion}\label{sec:conclusions}
In this paper, we propose a novel framework for deceptive jamming detection in monostatic \ac{ISAC}-\ac{OFDM} systems. This framework leverages the received signal at the \ac{BS} to detect the presence of a jammer capable of falsifying target localization.
The proposed framework employs a \ac{VAE} to learn a latent space representation of the echoes received from a target. Specifically, the reconstruction probability is utilized as a test statistic to detect the presence of a jammer.
Our approach demonstrates significant detection performance, achieving a detection probability $P_\text{d}$ of $93\%$ for an \ac{SJR} of $27\,$\text{dB}, and notably outperforming a properly trained conventional \ac{AE}.

\bibliographystyle{IEEEtran}
\bibliography{IEEEabrv,Biblio}  
\end{document}